\documentclass[prd,11pt,a4paper,nofootinbib,showkeys,tightenlines]{revtex4}
\usepackage[left=2.5cm,right=2.5cm,top=2.5cm,bottom=2.5cm]{geometry}

\usepackage[latin1]{inputenc}
\usepackage{enumerate}
\usepackage{amsmath}
\usepackage{amsthm}
\usepackage{amssymb}
\usepackage{graphicx}
\usepackage{floatflt}
\usepackage{fancybox}
\usepackage{appendix}
\usepackage{enumitem}
\usepackage{subfigure}
\usepackage{array}

\usepackage{hyperref}

\usepackage{color}

\newcommand{\om}{\omega}

\newcommand{\be} {\begin{equation}}
\newcommand{\ee} {\end{equation}}
\newcommand{\bsub}{\begin{subequations}}
\newcommand{\esub}{\end{subequations}}
\newcommand{\bea}{\begin{eqnarray}}
\newcommand{\eea}{\end{eqnarray}}
\newcommand{\bi} {\begin{itemize}}
\newcommand{\ei} {\end{itemize}}
\newcommand{\ben} {\begin{enumerate}}
\newcommand{\een} {\end{enumerate}}
\newcommand{\bmat} {\begin{pmatrix}}
\newcommand{\emat} {\end{pmatrix}}
\newcommand{\bal} {\begin{aligned}}
\newcommand{\eal} {\end{aligned}}
\newcommand{\btab}{\begin{tabular}}
\newcommand{\etab}{\end{tabular}}

\newcommand{\FigSize}{0.6\columnwidth}

\newcommand{\eq}[1]{Eq.~\eqref{#1}}

\begin{document}
%\selectlanguage{english}

\title{Subwavelength Su-Schrieffer-Heeger topological modes in acoustic waveguides}

\author{Antonin Coutant}
\email{antonin.coutant@ens-lyon.org}
\affiliation{Laboratoire d'Acoustique de l'Université du Mans (LAUM), UMR 6613, Institut d'Acoustique - Graduate School (IA-GS), CNRS, Avenue O. Messiaen, F-72085 Le Mans Cedex 9, France}
\affiliation{Institut de Math\' ematiques de Bourgogne (IMB), UMR 5584, CNRS, Universit\' e de Bourgogne Franche-Comt\' e, F-21000 Dijon, France}

\author{Vassos Achilleos} 
\email{achilleos.vassos@univ-lemans.fr}
\affiliation{Laboratoire d'Acoustique de l'Université du Mans (LAUM), UMR 6613, Institut d'Acoustique - Graduate School (IA-GS), CNRS, Avenue O. Messiaen, F-72085 Le Mans Cedex 9, France}

\author{Olivier Richoux}
\email{olivier.richoux@univ-lemans.fr}
\affiliation{Laboratoire d'Acoustique de l'Université du Mans (LAUM), UMR 6613, Institut d'Acoustique - Graduate School (IA-GS), CNRS, Avenue O. Messiaen, F-72085 Le Mans Cedex 9, France}

\author{Georgios Theocharis}
\email{georgios.theocharis@univ-lemans.fr}
\affiliation{Laboratoire d'Acoustique de l'Université du Mans (LAUM), UMR 6613, Institut d'Acoustique - Graduate School (IA-GS), CNRS, Avenue O. Messiaen, F-72085 Le Mans Cedex 9, France}

\author{Vincent Pagneux}
\email{vincent.pagneux@univ-lemans.fr}
\affiliation{Laboratoire d'Acoustique de l'Université du Mans (LAUM), UMR 6613, Institut d'Acoustique - Graduate School (IA-GS), CNRS, Avenue O. Messiaen, F-72085 Le Mans Cedex 9, France}

\date{\today}

\begin{abstract}
Topological systems furnish a powerful way of localizing wave energy at edges of a structured material. Usually this relies on Bragg scattering to obtain bandgaps with nontrivial topological structures. However, this limits their applicability to low frequencies since that would require very large structures. A standard approach to address the problem is to add resonating elements inside the material to open gaps in the subwavelength regime. Unfortunately, one usually has no precise control on the properties of the obtained topological modes, such as their frequency or localization length. In this work, we propose a new construction to couple acoustic resonators such that acoustic modes are mapped exactly to the eigenmodes of the Su-Schrieffer-Heeger model. The relation between energy in the lattice model and the acoustic frequency is controlled by the characteristics of the resonators. This allows us to obtain Su-Schrieffer-Heeger topological modes at any given frequency, for instance in the subwavelength regime. We also generalize the construction to obtain well-controlled topological edge modes in alternative tunable configurations. 
\end{abstract}


\maketitle

%\tableofcontents

%%%%%%%%%%%%%%%%%%%%%%%%%%%%%%%%%%%%%%%%%%%%%%%%%%%
%%%%%%%%%%%%%%%%%%%%%%%%%%%%%%%%%%%%%%%%%%%%%%%%%%%
%%%%%%%%%%%%%%%%%%%%%%%%%%%%%%%%%%%%%%%%%%%%%%%%%%%
%
%							INTRODUCTION
%
%%%%%%%%%%%%%%%%%%%%%%%%%%%%%%%%%%%%%%%%%%%%%%%%%%%
%%%%%%%%%%%%%%%%%%%%%%%%%%%%%%%%%%%%%%%%%%%%%%%%%%%
%%%%%%%%%%%%%%%%%%%%%%%%%%%%%%%%%%%%%%%%%%%%%%%%%%%
\section{Introduction}

The field of topological insulators has now found a large set of interests outside of electronic systems where it was first discovered, and it offers a powerful new approach for wave control in various contexts~\cite{Huber16,Zhang18,Ozawa19,Delplace20}. Among them, acoustic waves provide an ideal platform to implement and test the exotic properties of topological phases due to their high degree of tunability. For instance, several systems with non-trivial topology have been obtained based on appropriately chosen phononic crystal structures in one dimensions~\cite{Xiao14,Xiao15,Xiao17,Meng18} or higher~\cite{He16}. Such an approach usually relies on Bragg scattering, and hence, leads to structures that are several times larger than the typical wavelength, which can be rather large, especially in audible acoustics. 

To remedy this issue, various works have proposed to combine phononic crystal structures with subwavelength resonators~\cite{Yves17}. In particular, in one dimensional systems, several works have obtained topologically protected localized modes in the subwavelength regime using a chain of Helmholtz resonators~\cite{Li20,Zhao21}. The distances between pairs of resonators are then shifted, which opens a gap through a band folding mechanism and with distinct topology depending on whether the shift puts the resonators closer of further. A big limitation to this approach is that one has no control \emph{a priori} on the characteristics of the topological mode, such as its eigenfrequency or localization length. Moreover, the associated topological invariant (the Zak phase) is protected by mirror symmetry, and hence, is usually not robust to the introduction of spatial disorder. 

In this work, we propose an alternative approach based on an exact mapping to the Su-Schrieffer-Heeger (SSH) model recently developed~\cite{Coutant21}. We consider (subwavelength) resonators placed at the middle of segments of waveguides of equal length, but with cross sections that alternate between two values. We show that in the limit of narrow tubes, the acoustic modes of the system can be exactly mapped to the spectrum of the SSH model, known for its topological properties~\cite{Su79,Asboth16}. The characteristics of the resonators simply change the correspondence between SSH eigenvalues (the pseudo-energy E in the the following) and acoustic eigenfrequencies, which allows us to modify the frequency of the topologically protected edge modes at will. Similarly, the localization length of the topological mode is directly controlled by the cross-section change. In addition, edge modes of the SSH model are protected by chiral symmetry and characterized by a winding number, which means that the topological modes are robust to the introduction of symmetry preserving disorder. Again, through the exact mapping, we have a large range of frequencies where the acoustic waveguide mimics the discrete SSH model 
and we have a direct control on the allowed class of disorder against which the topological mode is robust~\cite{Coutant21}.  

This paper is organized as follows. In section~II, we introduce the general formalism to describe acoustic modes in varying cross-section waveguides with scatterers, and map it to the SSH model. In section~III we use Helmholtz resonators to obtain topologically protected modes localized near the edges of the system and we display direct finite element results confirming the existence of this kind of mode. 
In section~IV, we present an alternative configuration based on Helmholtz resonators put in series rather than in parallel. 

%%%%%%%%%%%%%%%%%%%%%%%%%%%%%%%%%%%%%%%%%%%%%%%%%%%
%%%%%%%%%%%%%%%%%%%%%%%%%%%%%%%%%%%%%%%%%%%%%%%%%%%
%%%%%%%%%%%%%%%%%%%%%%%%%%%%%%%%%%%%%%%%%%%%%%%%%%%
%
%							1D 		SSH		MODEL
%
%%%%%%%%%%%%%%%%%%%%%%%%%%%%%%%%%%%%%%%%%%%%%%%%%%%
%%%%%%%%%%%%%%%%%%%%%%%%%%%%%%%%%%%%%%%%%%%%%%%%%%%
%%%%%%%%%%%%%%%%%%%%%%%%%%%%%%%%%%%%%%%%%%%%%%%%%%%
\section{From acoustic waveguides to the Su-Schrieffer-Heeger model}
\label{SSH_Waveguide_Sec}

\begin{figure}[htp]
\centering
\includegraphics[width=\FigSize]{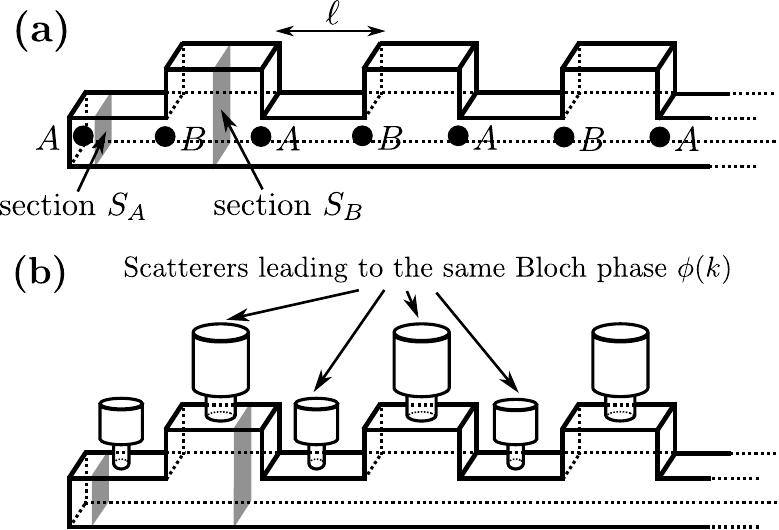}
\caption{(a) Representation of an acoustic waveguide with cross-section changes. Each cross section change is labelled with $j=\{A,B\}$ such that the cross section on the right side of $j$ is $S_j$. (b) Same as (a) but each segment contains a scatterer (indicated by a circle) inducing a nontrivial Bloch phase. 
}
\label{SSH_Waveguide_Fig} 
\end{figure}

\subsection{Reminder: SSH model in acoustic waveguides}
\label{CanSSH_acWG}
For the sake of clarity, we remind briefly the direct mapping approach proposed in~\cite{Coutant21}.
We consider the propagation of acoustic waves at fixed frequency $\om = k c_0$, with the harmonic convention $e^{-i \om t}$. $k$ is the wavenumber and $c_0$ the speed of sound. The waves propagate inside a waveguide obtained by connecting segments of alternating cross sections $S_A$ and $S_B$, as illustrated in Fig.~\ref{SSH_Waveguide_Fig}(a). All segments have the same length $\ell$. The waveguide is assumed to be sufficiently narrow so that propagation is one-dimensional (monomode propagation) along the $x$-axis. More precisely, we assume that the longitudinal length is much larger than the transverse ones. We describe the propagation using the transfer matrix formalism. By construction, the transfer matrix relates the acoustic pressure $p(x)$ and its derivative $p'(x)$ from one side to the other of a segment: 
\bsub \bea
\bmat p(x_n^B) \\ p'(x_n^B) \emat &=& M \cdot \bmat p(x_n^A) \\ p'(x_n^A) \emat , \\
\bmat p(x_{n+1}^A) \\ p'(x_{n+1}^A) \emat &=& M \cdot \bmat p(x_n^B) \\ p'(x_n^B) \emat , 
\eea \esub
where 
\be \label{Straight_Mmat}
M = \bmat \cos(k \ell) & \frac{\sin(k \ell)}{k} \\ -k \sin(k \ell) & \cos(k \ell) \emat . 
\ee
Now, to relate the propagation inside one segment to nearby ones, we use the continuity of pressure and acoustic flow rate at each cross section change. This gives the junction conditions  
\be
[p] = 0 \quad \text{and} \quad [S p'] = 0.
\ee
The idea is to start at a given cross section change, and relate the pressure there to the left and to the right pressures using the transfer matrix. Considering the section change at $x_n^A$, this leads to the pair of equations 
\bsub \label{P_eq1} \bea
p(x_{n}^B) &=& \cos(k \ell) p(x_{n}^A) + \frac{\sin(k \ell)}{k} p'_+(x_{n}^A) , \quad \label{forBC1} \\
p(x_{n-1}^B) &=& \cos(k \ell) p(x_{n}^A) - \frac{\sin(k \ell)}{k} p'_-(x_{n}^A) , \quad 
\eea \esub
where $p'_\pm(x)$ is short for the limit $\epsilon \to 0^\pm$ of $p'(x+\epsilon)$. 
Similarly, starting from $x=x_n^B$, we obtain 
\bsub \label{P_eq2} \bea
p(x_{n+1}^A) &=& \cos(k \ell) p(x_{n}^B) + \frac{\sin(k \ell)}{k} p'_+(x_{n}^B) , \quad \\
p(x_{n}^A) &=& \cos(k \ell) p(x_{n}^B) - \frac{\sin(k \ell)}{k} p'_-(x_{n}^B) . \quad \label{forBC2}
\eea \esub
Now, in both pairs of equations \eqref{P_eq1} and \eqref{P_eq2}, we can eliminate the pressure derivative term by using the flow rate continuity. This leads to an eigenvalue problem for the discrete set of pressure amplitudes $A_n \equiv p(x_n^A)$ and $B_n \equiv p(x_n^B)$: 
\bsub \label{Discrete_Model} \bea
\cos(k \ell) A_n &=& s B_n + t B_{n-1} , \\
\cos(k \ell) B_n &=& s A_n + t A_{n+1} , 
\eea \esub
with 
\bsub \bea
s &=& \frac{S_A}{S_A+S_B} , \\
t &=& \frac{S_B}{S_A+S_B} . 
\eea \esub
The eigenvalue problem obtained in equation~\ref{Discrete_Model} is \emph{exactly} the SSH model~\cite{Su79,Asboth16}. For finite waveguides, say made of an odd number $2N-1$ of segments and closed at both ends, the problem reduces to finding the eigenvalues of a $2N \times 2N$ matrix, analogous to the SSH Hamiltonian. At the closed ends the acoustic velocity vanishes, which imposes the condition $\cos(k \ell) A_1 = B_1$, and $\cos(k \ell) B_N = A_N$ (this can be directly obtained from equations \eqref{forBC1} and \eqref{forBC2}~\footnote{We refer the reader to \cite{Coutant21} section~III for an extended discussion on boundary conditions. In particular, it is explained how $H$ can be made hermitian by a rescaling of pressure values.}). Hence, the eigenvalue problem of \eq{Discrete_Model} has the form 
\be \label{eig_prob}
H  X = E X, 
\ee
with 
\be \label{SSH_Ham}
H = \bmat 0 & 1 & 0 & \dots & 0 \\ 
s & 0 & t & & \vdots \\ 
0 & t & 0 & \ddots & 0  \\ 
\vdots & & \ddots & \ddots & s \\ 
0 & \dots & 0 & 1 & 0 \emat ,
\ee
and $X = \bmat A_1, B_1, \dots A_N, B_N \emat^T$. The eigenvalue $E$ then gives the acoustic eigenfrequencies through the relation 
\be \label{Straight_EFrel}
E(k) = \cos(k \ell). 
\ee
The correspondance between waveguides of alternating cross sections was already established and studied in~\cite{Coutant21}, and also used to obtain two dimensional generalizations~\cite{Zheng19,Zheng20,Coutant20,Coutant20b}. In such a setup, properties of the SSH model are directly realized in acoustics. In particular, the SSH model is known to host a topologically protected mode at $E=0$, which, from \eq{Straight_EFrel}, induces topologically protected acoustic modes at $k \ell = \pm \pi/2+2n\pi$ with $n \in \mathbb N$. Unfortunately, this mapping limits the wavelength of theses modes to be at least of the order of a segment size. The novelty of the present work is to modify the energy-frequency relation of \eq{Straight_EFrel} %is that by tuning the Bloch phase $\phi(k)$, we can 
in order to obtain SSH-like modes in different frequency ranges. With a proper configuration, we now show that $k\ell$ in \eq{Straight_EFrel} is replaced  by a phase $\phi(k)$ that can be tailored to induce a topological mode with a  lower frequency.

\subsection{Generalized waveguides}

We now consider a broader class of waveguide generalizing section~\ref{CanSSH_acWG}, where each waveguide segment can host various scatterers, as shown in Fig.~\ref{SSH_Waveguide_Fig}(b). Propagation in each segment is given in terms of a transfer matrix: 
\bsub \bea
\bmat p(x^B) \\ p'(x^B) \emat &=& M_A \cdot \bmat p(x^A) \\ p'(x^A) \emat , \\
\bmat p(x^A) \\ p'(x^A) \emat &=& M_B \cdot \bmat p(x^B) \\ p'(x^B) \emat . 
\eea \esub
The key point of our generalization is to assume that the scattering induced is identical in each segment irrespectively of the cross section value, i.e. 
\be \label{Mmat_cond}
M_A = M_B . 
\ee
We also assume that the scattering is reciprocal, and each segment is mirror symmetric. Under these assumptions, the transfer matrix has the general form: 
\be \label{Mmat_def}
M = \bmat \alpha & \beta \\ \tilde \beta & \alpha \emat , 
\ee
with $\det(M)=1$. Note that the diagonal elements are equal due to mirror symmetry. As we shall see, the coefficient $\alpha$ plays a key role in our construction, and it is therefore useful to notice that it gives the dispersion relation of a medium obtained by connecting periodically segments described by $M$ (i.e. without cross section changes). Indeed, in this case solutions can be given in terms of Bloch waves satisfying $p(x_{n+1}^A) = e^{i \phi} p(x_{n}^A)$, with $\phi$ the Bloch phase. Since Bloch waves are also eigenvectors of the transfer matrix~\cite{Soukoulis}, the two eigenvalues are $e^{\pm i \phi}$. Since their sum is the trace of $M$, the dispersion relation is simply given by $\cos(\phi) = \alpha$. Hence, the transfer matrix can be written under a form very similar to \eq{Straight_Mmat} with the help of the Bloch phase $\phi$: 
\be \label{Gener_DispRel}
M = \bmat \cos(\phi) & \frac{\sin(\phi)}{Y} \\ -Y \sin(\phi) & \cos(\phi) \emat , 
\ee
where $Y$ is a complex number. Notice that by convention, we will now focus on $\phi>0$. We now follow exactly the same steps as in section~\ref{CanSSH_acWG}. We first write the propagation inside the segments 
\bsub \bea
p(x_{n}^B) &=& \alpha p(x_{n}^A) + \beta p'_+(x_{n}^A) , \\
p(x_{n-1}^B) &=& \alpha p(x_{n}^A) - \beta p'_-(x_{n}^A) ,  
\eea \esub
and
\bsub \bea
p(x_{n+1}^A) &=& \alpha p(x_{n}^B) + \beta p'_+(x_{n}^B) ,  \\
p(x_{n}^A) &=& \alpha p(x_{n}^B) - \beta p'_-(x_{n}^B) . 
\eea \esub
Again, using continuity of pressure and flow rate, one can get rid of pressure derivative in the preceding equations. Doing so, we obtain the exact same eigenvalue problem of \eq{eig_prob} with the Hamiltonian $H$ of \eq{SSH_Ham}. However, the energy eigenvalue is now related to the acoustic frequencies through the Bloch phase $\phi(k)$, hence, \eq{Straight_EFrel} is replaced by  
\be \label{EF_relation}
E(k) = \alpha(k) = \cos(\phi(k)). 
\ee
What we have shown is that for each eigenvalue $E$ of the SSH model \eqref{Discrete_Model} corresponds a set of eigenfrequencies of the waveguide obtained from \eq{EF_relation} controlled by the Bloch phase $\phi(k)$.

\subsection{Tuning the Bloch phase $\phi(k)$}

\begin{figure}[htp]
\centering
\includegraphics[width=\columnwidth]{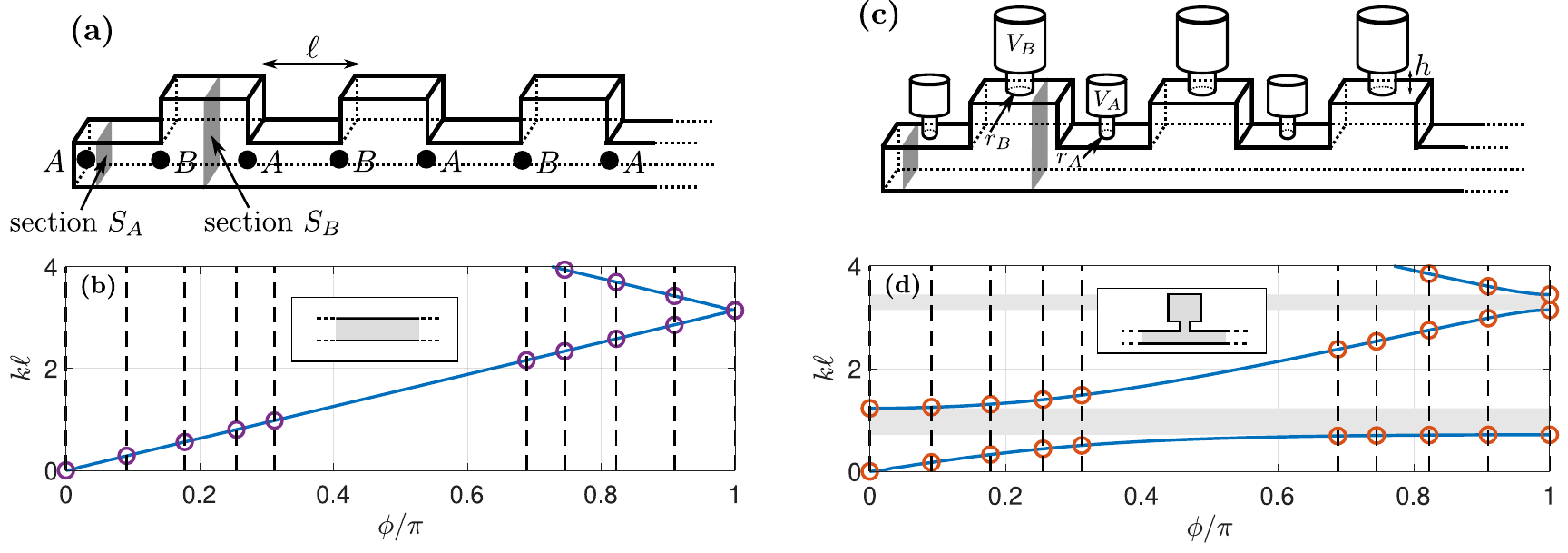} 
\caption{(a) Simple waveguide configuration with alternating cross-section where each segment is a straight tube. (b) Dispersion relation of a periodic arrangement of one of the two segments (shown in inset), i.e. $\phi(k) = k\ell \; \text{mod} \; 2\pi$. We then consider a cavity made of $9$ segments ($N=5$) and $S_A/S_B=3$: the dashed lines shows the values of $\phi$ giving associated with the Hamiltonian \eqref{SSH_Ham}, from which we obtain the acoustic eigenfrequencies (circles). (c) Representation of the acoustic waveguide with Helmoltz resonators at the middle of each segment. The changes of geometry between resonators $A$ and resonators $B$ is illustrated to scale. However, the neck radii are exaggerated to illustrate these changes, but they must both verify $r_A, r_B \ll \ell$. (d) Same as (b) for configuration with resonators shown in (c) ($N=5$ and $S_A/S_B=3$). 
}
\label{HR_Waveguide_Fig} 
\end{figure}

As mentioned before, when the waveguide segments are made of straight tubes with no scatterer, as in Fig.~\ref{HR_Waveguide_Fig}(a), the Bloch phase is simply given by $\phi(k) = k \ell$. Hence, to obtain the spectrum of a finite waveguide with cross-section changes, we look for frequencies such that $k\ell = \phi_n = \mathrm{acos}(E_n)$ with $E_n$ the eigenvalues of the SSH Hamiltonian of \eq{SSH_Ham}. This is illustrated in Fig.~\ref{HR_Waveguide_Fig}(b). In particular, since the SSH model has a gap around $E=0$, the cross-section changes induce the opening of a gap around $k\ell = \pi/2$. 

We now consider the waveguide configuration where a Helmholtz resonator is put on the upper wall at the middle of each segment, as shown in Fig.~\ref{HR_Waveguide_Fig}(c). We use a low frequency model for the Helmholtz resonators~\cite{Zheng20}, based on the following assumptions: (i) we only consider the lowest resonance frequency, (ii) we assume that the neck radius is much smaller than the typical wavelength (hence, $r \ll \ell$), (iii) we neglect dissipation. Under these assumptions, the resonators are characterized by two parameters: their resonance frequency $\om_0 = k_0 c_0$ and their coupling to the waveguide $g$. In the limit of a small neck, these parameters can be approximated by 
\bsub \label{HR_param_order1} \bea 
k_0 &=& \sqrt{\frac{\pi r^2}{V_c h}}, \\
g &=& \frac{\pi r^2}{S_0 h}, 
\eea \esub
where $V_c$ is the volume of the cavity, $r$ the radius of the neck, $h$ its length, and $S_0$ the cross section of the waveguide. 
Since the neck radius is much smaller than the typical wavelength, the resonators induce a jump of the acoustic velocity from one side and the other of the hole, while the pressure $p$ is continuous. This jump translates into a jump for the pressure derivative $[p'] = Y_0 p$ with 
\be \label{HRjump}
Y_0 = \frac{g k^2}{k^2 - k_0^2}. 
\ee
Now, to apply the results of section~\ref{SSH_Waveguide_Sec}, the transfer matrices of each segment must be equal as required in \eq{Mmat_cond}. For this, we need to make sure that all resonators have the same resonance frequency $k_0$ and the same coupling constant $g$. Since the latter depends on the section of the segment, we need to adjust the geometries of the resonators accordingly. In a regime where \eq{HR_param_order1} is valid, we might take $r_A^2/S_A = r_B^2/S_B$ and $r_A^2/V_A = r_B^2/V_B$. 

We can now compute the transfer matrix of a segment of length $\ell$ with a resonator at the middle point. Under the form of \eq{Mmat_def}, the matrix coefficients are given by 
\bsub \bea
\alpha &=& \cos(k\ell) + \frac{Y_0}{2k} \sin(k\ell) , \\
\beta &=& \frac1k \sin(k\ell) - \frac{Y_0}{2k^2} (\cos(k\ell)-1) , \\
\tilde \beta &=& -k \sin(k\ell) + \frac{Y_0}2 (\cos(k\ell)+1) . 
%M = \bmat \cos(k\ell) - z \sin(k\ell) & \frac1k \sin(k\ell) + \frac zk (\cos(k\ell)-1) \\ -k \sin(k\ell) - zk (\cos(k\ell)+1) & \cos(k\ell) - z \sin(k\ell) \emat . 
\eea \esub
As in the preceding section (\eq{Gener_DispRel}), the dispersion relation is directly obtained from the diagonal elements of $M$. This gives the Bloch phase as a function of the frequency $k$: 
\be \label{HR_BlochPhase}
\cos(\phi(k)) = \cos(k\ell) - \dfrac{g k}{2(k_0^2 - k^2)} \sin(k\ell). 
\ee
The acoustic eigenfrequencies of a finite waveguide can now be obtained by using the dispersion relation with the spectral values of the phase (see Fig.~\ref{HR_Waveguide_Fig}(c,d)).

%%%%%%%%%%%%%%%%%%%%%%%%%%%%%%%%%%%%%%%%%%%%%%%%%%%
%%%%%%%%%%%%%%%%%%%%%%%%%%%%%%%%%%%%%%%%%%%%%%%%%%%
%%%%%%%%%%%%%%%%%%%%%%%%%%%%%%%%%%%%%%%%%%%%%%%%%%%
%
%							TOPOLOGICAL MODE
%
%%%%%%%%%%%%%%%%%%%%%%%%%%%%%%%%%%%%%%%%%%%%%%%%%%%
%%%%%%%%%%%%%%%%%%%%%%%%%%%%%%%%%%%%%%%%%%%%%%%%%%%
%%%%%%%%%%%%%%%%%%%%%%%%%%%%%%%%%%%%%%%%%%%%%%%%%%%
\section{Topologically protected edge mode in the subwavelength regime}

We now show that the topological edge mode of the SSH model induces a topologically protected acoustic mode whose eigenfrequency can be controlled by the Bloch phase $\phi(k)$. 

\subsection{Reminder: Su-Schrieffer-Heeger edge modes}
\label{SSH_model_Sec}
\begin{figure}[htp]
\centering
\includegraphics[width=\FigSize]{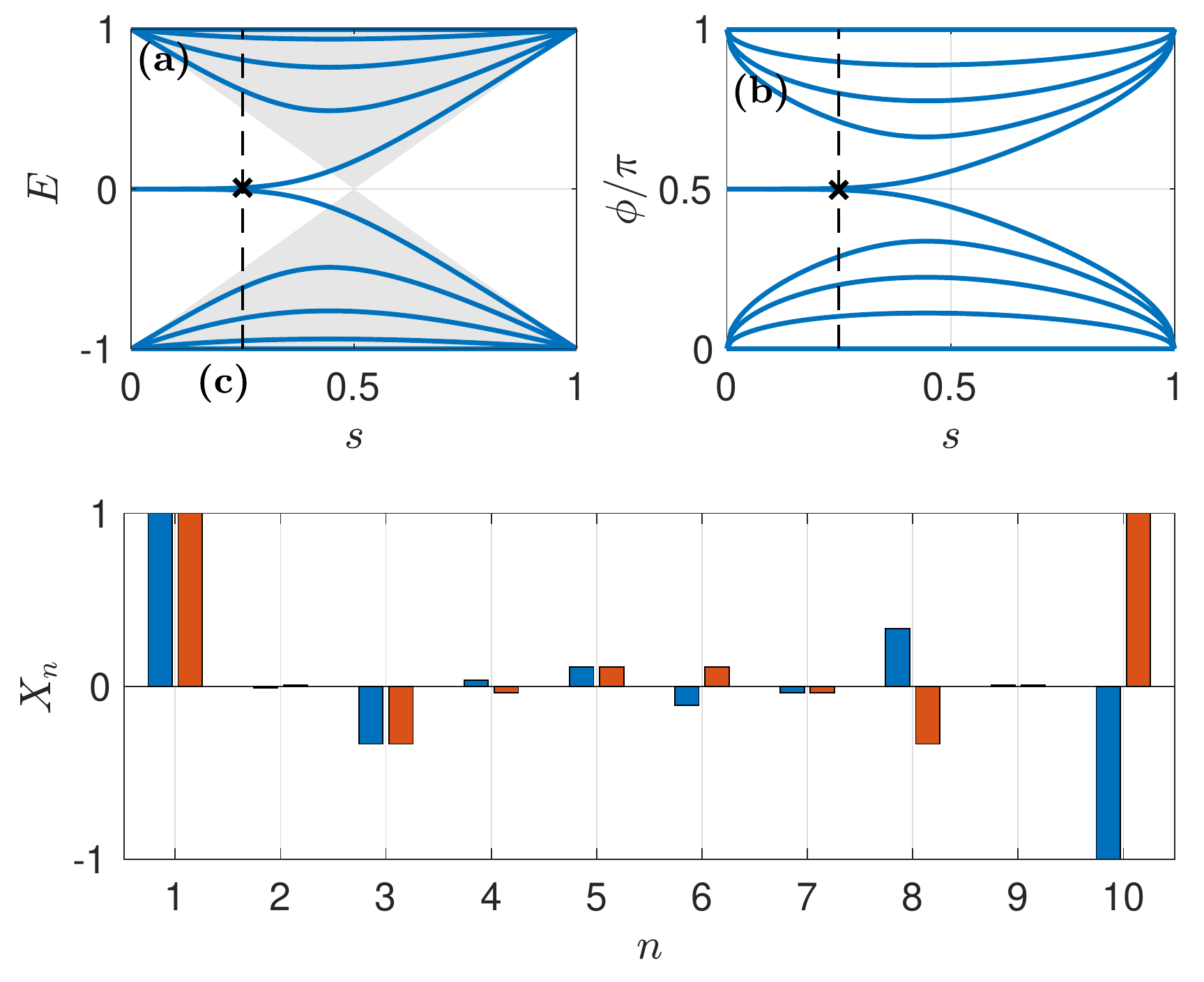} 
\caption{(a) Eigenvalues of an SSH chain with $2N=10$ sites for varying $s$ and $t=1-s$. The bands of the infinite chain are shown in light grey. (b) Bloch phase values corresponding to the spectrum (a). (c) Profil of the discrete edge modes $X_n$ for $s=0.25$ (marked by a black cross in (a) and (b)). 
}
\label{SSH_EtoPhi_Fig} 
\end{figure}

The SSH model is classical for its topological properties. In particular, the bulk-edge correspondance guarantees that when it is in its topological phase, edge modes are present in the middle of the gap, more specifically at $E=0$. This can be seen either by computing the appropriate topological invariant, the winding number or the Zak phase~\cite{Asboth16}, or by explicitly computing the edge mode. In particular, in the limit of a semi-infinite waveguide ($N\to \infty$) where the right end is sent to infinity, a direct verification shows that the mode 
\be \label{SSH_EdgeMode}
\bmat A_n \\ B_n \emat = -(s/t)^n \bmat 1 \\ 0 \emat 
\ee
is a solution of the eigenvalue problem~\eqref{Discrete_Model} with $E=0$. Moreover, it is localized on the left edge if $s<t$, which corresponds to the topological phase. In Fig.~\ref{SSH_EtoPhi_Fig}(a), we show the spectrum of the SSH model for varying parameter $s$ and $t=1-s$. As we just argued, for $t<s$ we only have extended modes in the passing bands (trivial phase), while for $s<t$, we have two edge modes localized on each end of the chain (topological phase). From relation~\eqref{EF_relation}, for each of the eigenvalues $E_n$ corresponds a value of the Bloch phase $\phi_n>0$ and \emph{vice-versa}. This is shown in Fig.~\ref{SSH_EtoPhi_Fig}(b). One can then obtain the frequency spectrum of the acoustic system from the dispersion relation~\eqref{Gener_DispRel}. In particular, when $s<t$, we will obtain two topologically protected edge modes (one is symmetric and the other antisymmetric) for two frequencies corresponding to $\phi \simeq \pi/2$, as we see in Fig.~\ref{SSH_EtoPhi_Fig}(b).

\subsection{Subwavelength topological edge modes}
\label{HR_Topo_Sec}

\begin{figure}[htp]
\centering
\includegraphics[width=\FigSize]{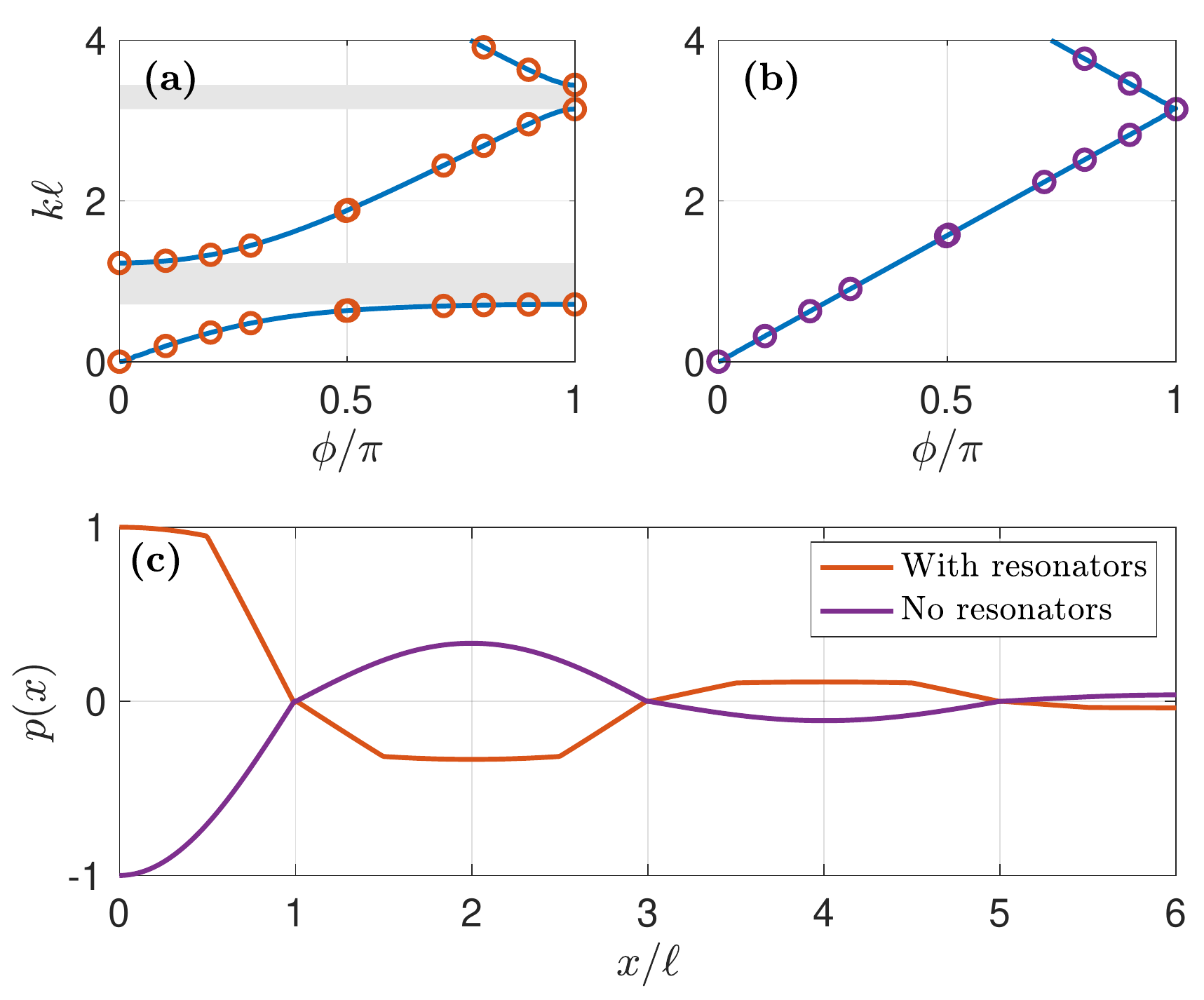}  
\caption{(a) Dispersion relation (blue line): frequency as a function of the Bloch phase (equation~\eqref{HR_BlochPhase}). We took $k_0 \ell =0.8$, $g \ell = 1$, and $S_A/S_B=1/3$. Spectrum (red circles) for $9$ segments with open ends. Gaps are shown in light grey. (b) Dispersion relation (blue line) and spectrum (purple circles) for $9$ segments in the absence of Helmholtz resonators and for $S_A/S_B=1/3$. (c) Pressure profiles of the left edge mode for both cases (a) and (b). 
}
\label{EdgeModes_Fig} 
\end{figure}

Turning back to the configuration of Fig.~\ref{HR_Waveguide_Fig}(c), the zero energy mode of the SSH model (seen in equation~\eqref{SSH_EdgeMode}) induces a localized topological acoustic mode each time $\phi(k) = \pi/2$. This is illustrated in Fig.~\ref{EdgeModes_Fig}(a). The isolated eigenfrequencies we see near $\phi=\pi/2$ (when $s<t$) correspond to edge modes. We compare this configuration with a waveguide of alternating cross sections and without resonators, shown in Fig.~\ref{HR_Waveguide_Fig}(c). We see that the presence of Helmholtz resonators opens a gap near the resonance frequency $k_0$, which in turn, induces extra edge modes at low frequencies $k<k_0$. This means that, by taking $k_0$ very small, we can obtain topological acoustic modes at arbitrary low frequencies. This is our main result: by changing the scatterer inside each segment, the mapping to the SSH model is exact and realized for any desired frequency range through \eq{EF_relation}. In Fig.~\ref{EdgeModes_Fig}(c), we compare the pressure profile of the edge mode on the left for waveguides with (red curve) or without resonators (purple curve).

\begin{figure}[htp]
\centering
\includegraphics[width=0.9\columnwidth]{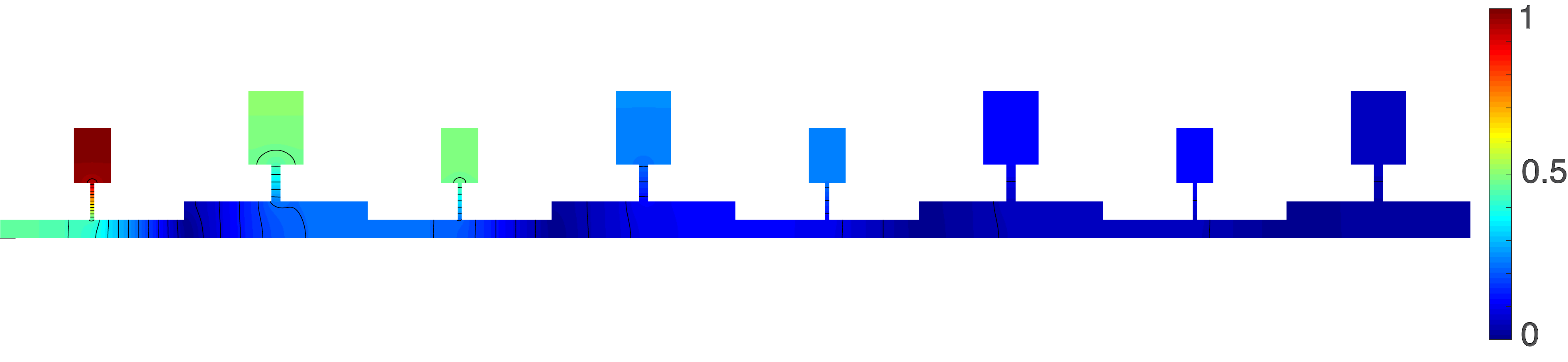}  
\caption{Numerical computation of the edge mode in a 2D geometry with resonators for $S_A/S_B=0.5$ (and thus $s=1/3$). The corresponding eigenfrequency is $k \ell = 0.30 \pi$. }
\label{EdgeModes_pdetool} 
\end{figure}

To confirm this subwavelength topological edge wave we have performed the numerical computation of eigenfrequencies and eigenmodes of
a cavity closed by hard walls. It solves the 2D Helmholtz equation $\triangle p + k^2 p$ with Neumann boundary conditions at the walls, using
Finite Element Method. We follow the approach described previously, but take an even number of segments ($N=8$), which allows us to isolate
one unique edge mode~\cite{Coutant21} at the extremity with the smallest width (here $S_A$).
Each segment is decorated by an Helmholtz resonator with neck width $r$, neck length $h$ and a cavity of 
width $L_x$ and height $L_y$. These geometrical parameters are finely tuned to get an equal Bloch phase following equation~\eqref{HR_BlochPhase} in the range of frequencies $0 \le k \ell \le \pi$, for both segments of cross section $S_A$ and $S_B$. The values of these tuned paramters are found to be $S_A=0.1 \ell$, $r_A=0.0217 \ell$,  $h_A=0.2 \ell$,  $L_{xA}=0.2 \ell$, $L_{yA}=0.3 \ell$  and $S_B=0.2 \ell$, $r_B=0.05 \ell$,  $h_B=0.2 \ell$,  $L_{xB}=0.3 \ell$, $L_{yB}=0.4 \ell$; they correspond to resonance frequency $k_0 \ell =1.226$ and a coupling parameter $g \ell =0.914$. 
The corresponding edge mode is confirmed to appear (it is the fifth mode of the cavity) and is displayed in Fig.~\ref{EdgeModes_pdetool}. It can be remarked that the amplitude is strongly concentrated in the first
resonator at the left.

\section{Remark: resonators in series by inverse design}

Interestingly, now that we understood the structure of the subwavelength edge mode (shown in Fig.~\ref{EdgeModes_Fig}(c)), we can reverse engineer it to obtain alternative structures also hosting a subwavelength edge mode. 

We now show how this can be done in an array of only waveguide segments of changing cross-section. First, we assume that we work at low frequencies, such that the pressure profile is piecewise linear (as in Fig.~\ref{EdgeModes_Fig}(c)). Let us start at the left wall, where the pressure derivative vanishes. Because of that boundary condition, the pressure profile in the first segment, of section $S_1$ is flat. We then put a segment of section $S_2 \ll S_1$, so that the pressure derivative at $x=\ell$ jumps from a near-zero value to a non-zero one. Since the profile in the first segment is not exactly flat, a small shift of frequency changes the near-zero derivative at $x=\ell$. The edge mode is such that the pressure slope in the second segment leads to a vanishing pressure at $x=2\ell$. Here, we put another segment $S_3>S_2$, so that the pressure derivative decreases in amplitude. Now, the fourth segment has a section satisfying $S_4 \gg S_3$, so that the pressure derivative becomes almost zero in that segment. Doing so, at the fifth section change, pressure is maximum and hence no jump of derivative occur, and we can put a segment of section $S_1$ again. However, we need to flip the sign of the (near-zero) derivative, and hence, we must use $S_1\lesssim S_4$. Doing so, the profile is back to its initial structure with an overall decrease of amplitude. By repeating the construction, we see that the pressure amplitude decreases for each set of four segment, and hence we obtain an edge mode, which by construction has a low (subwavelength) frequency. 

\begin{figure}[htp]
\centering
\includegraphics[width=\FigSize]{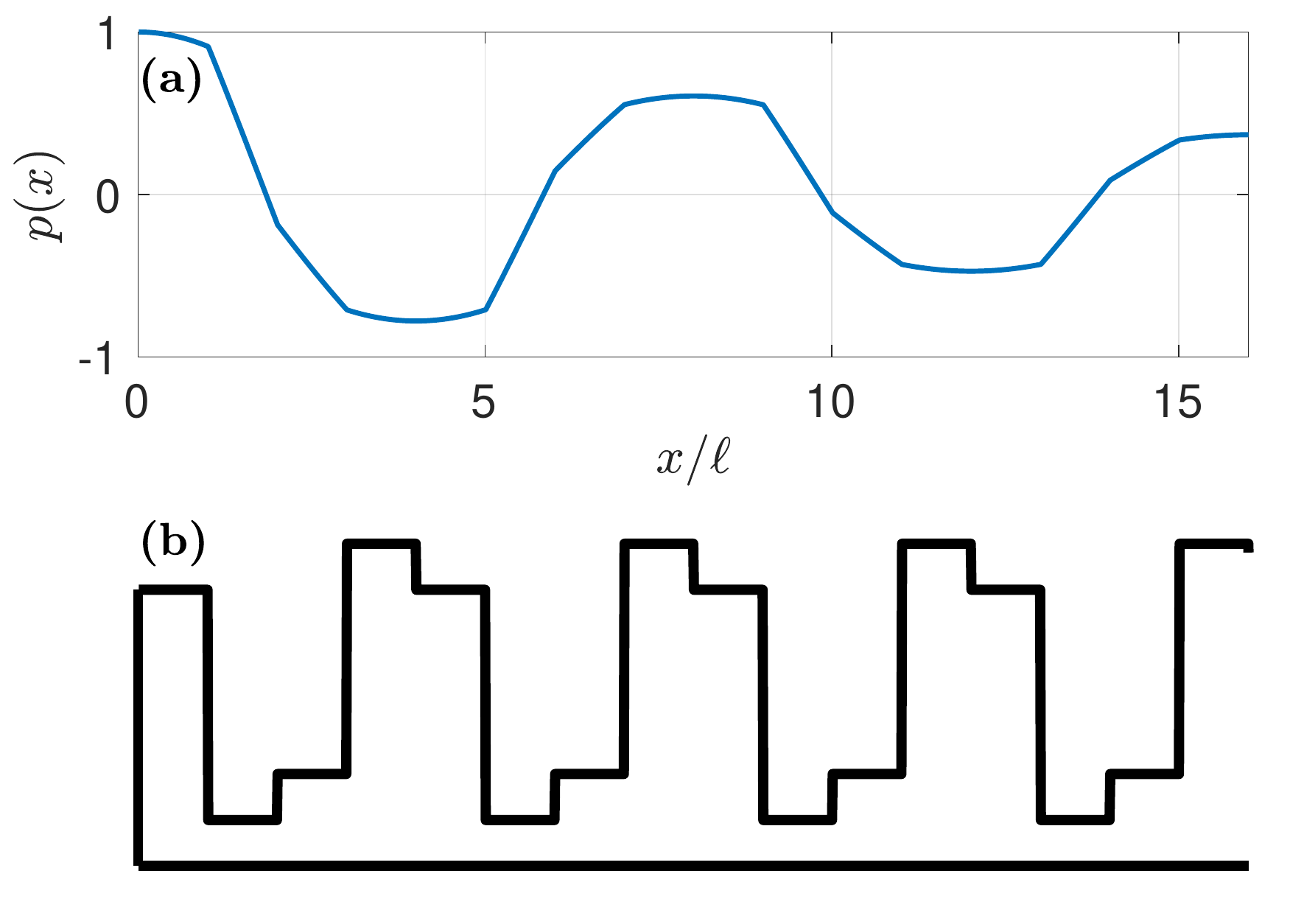} 
\caption{(a) Pressure profile of the lowest frequency edge mode (in blue) in the configuration of panel (b). (b) Waveguide configuration with repeating sets of four segments. }
\label{EdgeMode_SeriesRes}
\end{figure}

Because a picture is worth a thousand words, we show the edge mode in such a configuration in Fig.~\ref{EdgeMode_SeriesRes}. Notice that each large change of cross-section can be seen as a Helmholtz resonator. Hence, the described configuration amonts to Helmholtz resonators in placed in series, and inducing an edge mode near their resonance frequency.

Remarkably, while we have built a construction made of a set of four segments repeated along the waveguide, the same construction can be generalized to any values of cross-sections (e.g. in disordered configurations) as long as the inequalities exposed above are still valid. Explicitly, if we call $S_n$ the section of the $n^{\rm th}$ segment, we need $S_{4j+2} \ll S_{4j+1}$, $S_{4j+3}>S_{4j+2}$, $S_{4j+4} \gg S_{4j+3}$, and $S_{4j+5} \lesssim S_{4j+4}$. For completeness, we now give the discrete set of equation relating pressure at each cross-section change (more details for the derivation can be found in \cite{Coutant21}, section IV): 
\begin{widetext}
\bsub \bea
\cos(k\ell) p_{4j+1} &=& \frac{S_{4j}}{S_{4j}+S_{4j+1}} p_{4j} +  \frac{S_{4j+1}}{S_{4j}+S_{4j+1}} p_{4j+2} , \\
\cos(k\ell) p_{4j+2} &=& \frac{S_{4j+1}}{S_{4j+1}+S_{4j+2}} p_{4j+1} + \frac{S_{4j+2}}{S_{4j+1}+S_{4j+2}} p_{4j+3} , \\
\cos(k\ell) p_{4j+3} &=& \frac{S_{4j+2}}{S_{4j+2}+S_{4j+3}} p_{4j+2} + \frac{S_{4j+3}}{S_{4j+2}+S_{4j+3}} p_{4j+4} , \\
\cos(k\ell) p_{4j+4} &=& \frac{S_{4j+3}}{S_{4j+3}+S_{4j+4}} p_{4j+3} + \frac{S_{4j+4}}{S_{4j+3}+S_{4j+4}} p_{4j+5} , 
\eea \esub
\end{widetext}

\section{Conclusion}

In a previous work it was shown that a one-dimensional acoustic waveguide with alternating cross sections can be mapped exactly to the SSH lattice model~\cite{Coutant21}. Here, we extend this mapping by adding extra scatterers inside each segment of the waveguide. We show that the mapping is governed by a relation between the energy eigenvalue of the lattice model and the acoustic frequency, which now involves the Bloch phase accumulated along a segment (equivalently, the acoustic path), see equation~\ref{EF_relation}. Using this extended relation, we consider a configuration where Helmholtz resonators are added on the wall of each segment. Near the resonance frequency, it is now possible to obtain large Bloch phases. As a consequence, the topological edge mode of the SSH model, that has zero energy, is mapped to an acoustic mode whose frequency can be tuned at will. In particular, we obtain topologically protected modes at an arbitrarily low frequency. 

Moreover, our setup offers a much larger degree of control of the mode properties than previous proposals of a subwavelength topological mode~\cite{Li20,Zhao21}. Indeed, based on the exact mapping to the SSH model, the cross section changes and the Helmholtz resonators parameters govern distinct and well identified aspects of the system. For instance, the eigenfrequency of the edge mode is entirely controlled by the resonators characteristics, but insensitive to the cross section values, which in turn can be used to change the size of the corresponding gap.

%\clearpage
%\bibliographystyle{utphys}
\bibliography{Bibli}

\end{document}